\documentstyle[sprocl]{article}

\input{psfig.sty}

\def\be{\begin{equation}}
\def\ee{\end{equation}}
\def\bea{\begin{eqnarray}}
\def\eea{\end{eqnarray}}

\begin{document}

\title{Effects of $A_t$ and $\mu$ phases on $B$ and $K$ physics \\ in the 
effective supersymmetric models  
}

\author{S. BAEK, ~P. KO$^a$}

\address{Dep. of Physics, KAIST\\
Taejon 305-701, KOREA \\E-mail: $^a$ pko@muon.kaist.ac.kr} 




\maketitle\abstracts{ 
In the minimal supersymmetric standard model (MSSM), the $\mu-$parameter 
and the trilinear coupling $A_t$ may be generically complex and can affect 
various observables at B factories.  Working in the effective SUSY models and
imposing the edm constraints from Chang-Keung-Pilaftsis (CKP) mechanism, 
we find that (i) there is no new large phase shift in the 
$B^0 - \overline{B^0}$ mixing from the $A_t$ and $\mu$ phases, 
(ii) CP violating dilepton asymmetry is smaller than $0.1 \%$,   
(iii) the direct CP violation in $B\rightarrow X_s \gamma$ can be as 
large as $\sim \pm 16 \%$,
(iv) the ${\rm Br} (B\rightarrow X_s l^+ l^-)$ can be 
enhanced by upto $\sim 85 \%$ compared to the standard model (SM) prediction, 
and its correlation with ${\rm Br} ( B \rightarrow X_s \gamma)$ is distinctly 
different from the minimal supergravity scenario. Also, we find $1 \leq 
\epsilon_K / \epsilon_K^{SM} \leq 1.4$, and $\epsilon_K$ cannot be saturated
by the $A_t$ and $\mu$ phases alone : 
namely, $|\epsilon_K^{\rm SUSY}| \leq O(10^{-5})$ if the phases of $\mu$
and $A_t$ are the sole origin of CP violation.
}

\section{Introduction}
\subsection{Motivations}\label{subsec:prod}
In the minimal supersymmetric standard model (MSSM), there can be many new
CP violating (CPV) phases beyond the KM phase in the standard model (SM). 
These SUSY CPV phases are constrained by electron/neutron electric dipole 
moment (EDM) and have been considered very small ($\delta \leq 
10^{-2}$ for $M_{\rm SUSY} \sim O(100)$ GeV ). \cite{susycp} 
However there is a logical possibility that various contributions to 
electron/neutron EDM  cancel with each other in substantial part of the 
MSSM parameter space even if SUSY CPV phases are  $\sim O(1)$. 
\cite{nath}, \cite{kane}  Or one can consider effective  SUSY models
where decouplings of the 1st/2nd generation sfermions are invoked to evade
the EDM constraints and also SUSY FCNC/CP problems. \cite{kaplan} 
In such cases, these new SUSY phases may affect $B$ and $K$ physics 
in various manners. 
Closely related with this is the  electroweak baryogenesis (EWBGEN) 
scenario in the MSSM.
One of the fundamental problems in particle physics is to understand the
baryon number asymmetry, $n_B/s = 4 \times 10^{-12}$, and currently popular 
scenario is EWBGEN in the MSSM. \cite{carena}
The EWBGEN is in fact possible in a certain region of the MSSM parameter 
space, especially for light stop ($120~{\rm GeV} \leq m_{\tilde{t}_1} 
\leq 175$ GeV,  dominantly $\tilde{t}_1 \simeq \tilde{t}_R$) 
with CP violating phases in $\mu$ and $A_t$ parameters.
Then one would expect this light stop and new CP violating phases may  
lead to observable consequences to $B$ physics.

In this talk, we report our two recent works related with this  subject.
\cite{ko1},\cite{ko2}  
We considered a possibility of observing effects 
of these new flavor conserving and CPV phases ($\phi_\mu$ and $\phi_{A_t}$) 
at $B$ factories in the MSSM (including EWBGEN scenario therein).
More specifically, we consider the following observables : 
SUSY contributions to the $B^0 - \overline{B^0}$ mixing, the dilepton CP 
asymmetry in the $B^0 \overline{B^0}$ decays, the direct CP asymmetry 
in $B \rightarrow X_s \gamma$, the branching ratio for 
$B\rightarrow X_s l^+ l^-$ and its correlation with the branching ratio for
$B \rightarrow X_s \gamma$. The $B^0 - \overline{B^0}$ mixing is important 
for determination of three angles of the unitarity triangle. Also, 
last two observables are vanishingly small in the standard model (SM), 
and any appreciable amounts of these asymmetries would herald the existence 
of new CP violating phases beyond the KM phase in the SM. 
The questions addressed in our  papers were how much these observables can 
be deviated from their SM values when $\mu$ and $A_t$ parameters in the 
MSSM have new CPV phases.

\subsection{Model}
In order to study $B$ physics in the MSSM, we make the following assumptions. 
\cite{misiak}  First of all, the 1st and the 2nd family squarks are assumed 
to be degenerate and very heavy in order to solve the SUSY FCNC/CP problems.
\cite{kaplan}  
Only the third family squarks can be light enough to affect
$B\rightarrow X_s \gamma$ and $B^0 - \overline{B^0}$ mixing.
We also ignore possible flavor changing squark mass matrix elements
that could generate gluino-mediated flavor changing neutral current (FCNC)
process in addition to those effects we consider below. Recently, such 
effects were studied in the $B^0 - \overline{B^0}$ mixing, \cite{cohen-B}, 
\cite{randall} the branching ratio of $B\rightarrow X_s \gamma$ 
\cite{cohen-B} and CP violations therein, \cite{hou} \cite{kkl} and 
$B\rightarrow X_s l^+ l^-$, \cite{kkl} respectively. 
Ignoring such contributions, the only source of the FCNC in our model is the 
CKM matrix, whereas there are new CPV phases coming from the phases of $\mu$ 
and $A_t$ parameters in the flavor preserving sector in addition to the KM 
phase $\delta_{KM}$ in the flavor changing sector. In this sense, this paper
is complementary to the ealier works. \cite{cohen-B},\cite{randall},
\cite{hou},\cite{kkl}

\subsection{Chang-Keung-Pilaftsis (CKP) EDM Constraints}
Even if the 1st/2nd generation squarks are very heavy and degenerate, there
is another important edm constraints considered by Chang, Keung and Pilaftsis 
(CKP)  for large $\tan\beta$. \cite{pilaftsis}  
This constraint comes from the two loop diagrams involving stop/sbottom 
loops, and is independent of the masses of the 1st/2nd generation squarks.
\begin{equation}
( {d_f \over e } )_{\rm CKP} = Q_f {3 \alpha_{\rm em} \over 64 \pi^2}
{R_f  m_f \over M_A^2} \sum_{q=t,b} \xi_q Q_q^2 
F \left( {M_{\tilde{q}_1}^2 \over M_A^2}, {M_{\tilde{q}_2}^2 \over 
M_A^2 } \right)  
\end{equation} 
where $R_f = \cot\beta ~ (\tan\beta)$ for $I_{3 f} = 1/2 ~(-1/2)$, and 
\begin{equation}
\xi_t = {\sin 2\theta_{\tilde{t}} m_t {\rm Im} (\mu e^{i \delta_t} ) 
\over \sin^2 \beta ~v^2}, ~~~~
\xi_b = {\sin 2\theta_{\tilde{b}} m_b {\rm Im} ( A_b e^{-i \delta_b} ) 
\over \sin \beta ~\cos\beta~v^2},
\end{equation}
with $\delta_q = {\rm Arg} (A_q + R_q \mu^* )$, and $F(x,y)$ 
is a two-loop function given in Ref.~\cite{pilaftsis}.
Therefore, this CKP edm constraints can not be simply evaded by making the 
1st/2nd generation squarks very heavy, and it turns out that this puts a 
very strong constraint on the possible new phase shift in the 
$B^0 - \overline{B^0}$ mixing.  

\subsection{Parameter Space}
In the MSSM, the chargino mass matrix is given by
\begin{equation}
M_{\chi^{\pm}} = \left( 
\begin{array}{cc}
M_2 & \sqrt{2} m_W \sin\beta
\\
\sqrt{2} m_W \cos\beta  & \mu 
\end{array}
\right) .
\end{equation}
In principle, both $M_2$ and $\mu$ may be complex, but one can perform a phase 
redefinition in order to render the $M_2$ is real. \cite{kane} 
In such a basis, there appears one new phase ${\rm Arg} (\mu)$ 
as a new source of CPV. The stop mass matrix is given by 
\begin{equation}
M_{\tilde{t}}^2 = \left( 
\begin{array}{cc}
m_Q^2 + m_t^2 + D_L   & m_t ( A_t^* - \mu / \tan \beta ) 
\\
m_t ( A_t - \mu^* / \tan \beta ) & m_U^2 + m_t^2 + D_R
\end{array}
\right) ,
\end{equation}
where $D_{L} = ( {1\over 2} - {2\over 3}~ \sin^2 \theta_W ) \cos 2\beta ~
m_Z^2$ and $D_{R} =  {2\over 3} \sin^2 \theta_W \cos 2\beta ~m_Z^2$. 
There are two new phases in this matrix, ${\rm Arg} (\mu)$ and ${\rm Arg} 
(A_t)$ in the basis where $M_2$ is real. 

We scan over the MSSM parameter space as indicated below 
(including that relevant to the EWBGEN scenario in the MSSM) :
$  
80~{\rm GeV} < | \mu | < 1~{\rm TeV},~ 
80~{\rm GeV} < M_2 < 1~{\rm TeV},~
60~{\rm GeV} < M_A < 1~{\rm TeV},~
2 < \tan\beta < 70,~ 
(130~{\rm GeV})^2 <  M_Q^2  <  ( 1 ~{\rm TeV} )^2,~
  - ( 80~{\rm GeV})^2  <  M_U^2  <  (500~{\rm GeV})^2,~
0 < \phi_{\mu}, \phi_{A_t} < 2 \pi,~  0 < | A_t | < 1.5 
~{\rm TeV}.
$
We have imposed the following experimental constraints : 
$M_{\tilde{t}_1} > 80$ GeV independent of the mixing angle 
$\theta_{\tilde{t}}$, $M_{\tilde{\chi^{\pm}}} > 83$ GeV, 
${\rm Br} (B\rightarrow X_{sg}) < 6.8 \%$, \cite{bsg}  and
$0.77 \leq R_{\gamma} \leq 1.15$, \cite{alexander} 
where $R_{\gamma}$ is defined as 
$R_{\gamma}= BR(B \to X_s \gamma)^{expt}/ BR(B \to X_s \gamma)^{SM}$ and 
$BR(B \to X_s \gamma)^{SM} = (3.29 \pm 0.44) \times 10^{-4}$. 
It has to be emphasized that this parameter space is larger than that in  
the constrained MSSM (CMSSM) where the universality of soft terms at the GUT 
scale is assumed.  Especially, we will allow $m_U^2$ to be negative as well 
as positive,  which is preferred in the EWBGEN scenario. \cite{carena} 
Since we do not impose any further requirement on the soft 
terms (such as radiative electroweak symmetry breaking, absence of color 
charge breaking minima, etc.), our results of the maximal deviations of 
$B^0-\overline{B^0}$ mixing and $A_{\rm CP}^{b\rightarrow s\gamma}$
from the SM predictions are conservative upper bounds within the MSSM. 
If more theoretical conditions are imposed, the maximal deviations will 
be smaller. 
In the numerical analysis, we used the following numbers for the input 
parameters : $\overline{m_c}(m_c(pole)) = 1.25$ GeV, 
$\overline{m_b}(m_b(pole)) = 4.3$ GeV, $\overline{m_t}(m_t(pole)) = 165$ 
GeV (these are running masses in the $\overline{MS}$ scheme), 
and $| V_{cb} | = 0.0410, |V_{tb}| = 1, | V_{ts} | = 0.0400$ 
and $\delta_{KM} = \gamma (\phi_3) = 90^{\circ}$ for the CKM matrix elements.

\section{$B^0 - \overline{B^0}$ Mixing}
\subsection{Phase Shift in the $B^0 - \overline{B^0}$ Mixing}
The $B^0 - \overline{B^0}$ mixing is generated by the box diagrams 
with $u_i-W^{\pm} (H^{\pm})$ and $\tilde{u}_i-\chi^{\pm}$ running around 
the loops in addition to the SM contribution. The resulting effective 
Hamiltonian is given by 
\begin{equation}
H_{\rm eff}^{\Delta B=2} = - {G_F^2 M_W^2 \over (2\pi)^2}~
\sum_{i=1}^3 C_i O_i,
\end{equation}
where $O_1 = \overline{d}_L^{\alpha} \gamma_{\mu} b_{L}^{\alpha}~
\overline{d}_L^{\beta} \gamma^{\mu} b_{L}^{\beta},
O_2 = \overline{d}_{L}^{\alpha} b_{R}^{\alpha} \overline{d}_L^{\beta} 
b_{R}^{\beta}$, and  
$O_3 = \overline{d}_{L}^{\alpha} b_{R}^{\beta} \overline{d}_L^{\beta} 
b_{R}^{\alpha}$. 
The Wilson coefficients $C_i$'s at the electroweak scale ($\mu_0 \sim M_W 
\sim M_{\tilde{t}}$) can be written  schematically as \cite{branco}
\begin{eqnarray}
C_1 ( \mu_0 ) & = & \left( V_{td}^* V_{tb} \right)^2 ~
\left[ F_V^W (3;3) + F_V^H (3;3) +  A_V^C \right]
\nonumber  \\
C_2 ( \mu_0 ) & = & \left( V_{td}^* V_{tb} \right)^2 ~F_S^H (3;3)
\nonumber  \\
C_3 ( \mu_0 ) & = & \left( V_{td}^* V_{tb} \right)^2 ~A_S^C,
\end{eqnarray}  
where the superscripts $W,H,C$ denote the $W^{\pm}, H^{\pm}$ and chargino
contributions respectively, and  
\begin{eqnarray}
A_V^C  & = & \sum_{i,j,k,l}^{1,2} 
{1\over 4}~G_{(3,k)}^{i} G_{(3,k)}^{j*} G_{(3,l)}^{i*} G_{(3,l)}^{j} 
Y_1 (r_k, r_l, s_i, s_j ),
\nonumber  
\\
A_S^C & = & \sum_{i,j,k,l}^{1,2} 
H_{(3,k)}^{i} G_{(3,k)}^{j*} G_{(3,l)}^{i*} H_{(3,l)}^{j} 
Y_2 (r_k, r_l, s_i, s_j ),
\nonumber  
\end{eqnarray}
Here $G_{(3,k)}^{i}$ and $H_{(3,k)}^{i}$ are the couplings of $k-$th stop 
and $i-$th chargino with left-handed and right-handed quarks, respectively :
\begin{eqnarray}
G_{(3,k)}^{i} & = & \sqrt{2} C_{R 1i}^* S_{t k1} - 
{ C_{R 2i}^* S_{t k2} \over \sin\beta } ~{m_t \over M_W},
\nonumber \\
H_{(3,k)}^{i} & = & {C_{L 2 i}^* S_{tk1} \over \cos\beta } ~{m_b \over M_W},
\end{eqnarray}
and $C_{L,R}$ and $S_t$ are unitary matrices that diagonalize the chargino
and stop mass matrices : $C_R^{\dagger} M_{\chi}^- C_L = {\rm diag} 
( M_{\tilde{\chi_1}},M_{\tilde{\chi_2}} )$ and $S_t M_{\tilde{t}}^2
S_t^{\dagger} = {\rm diag} ( M_{\tilde{t}_1}^2, M_{\tilde{t}_2}^2 )$. 
Explicit forms for functions $Y_{1,2}$ and $F$'s can be found in Ref.~
\cite{branco}, and $r_k = M_{\tilde{t}_k}^2 / M_W^2$ and 
$s_i = M_{\tilde{\chi_i^{\pm}}} / M_W^2$. 
It should be noted that $C_2 ( \mu_0 )$ was misidentified as 
$C_3^H ( \mu_0 )$ in Ref.~\cite{demir}.
The gluino and neutralino contributions are negligible in our model.
The Wilson coefficients at the $m_b$ scale are obtained by renomalization
group running. The relevant formulae with NLO QCD corrections at $\mu = 2$ 
GeV are given in Ref.~\cite{contino}.

\begin{figure}
    \begin{center}
      \begin{picture}(140,200)
        \put(-98,-30){\psfig{file=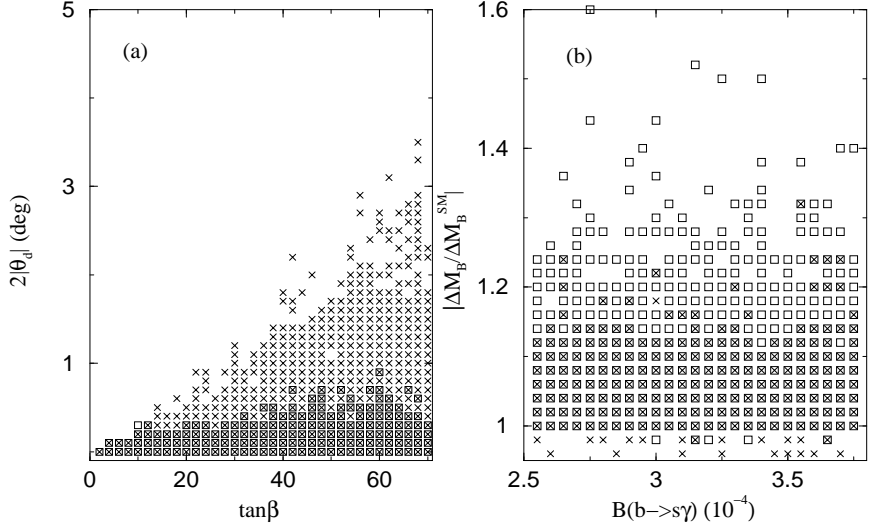,width=11.5cm}}
        \end{picture}
     \end{center}
\caption{
Correlations between
(a) $\tan\beta$ vs. 2 $|\theta_d | $, and 
(b) ${\rm Br} (B \rightarrow X_s \gamma)$ vs. 
$A_{12}^{\rm FULL} / A_{12}^{\rm SM}$. 
The squares (the crosses) denote those which (do not) satisfy the CKP edm 
constraints.  
}
\label{fig1}
\end{figure}

In our model, $C_1 ( \mu_0 )$ and $C_2 ( \mu_0 )$ are real  relative to the 
SM contribution. On the other hand, the chargino exchange contributions to 
$C_3 (\mu_0 )$ (namely $A_S^C $)are generically complex relative to the 
SM contributions, and  can generate a new phase shift in the 
$B^0 - \overline{B^0}$ mixing relative to the SM value. This effect can be 
in fact significant for large $\tan\beta (\simeq 1/\cos\beta)$, since 
$C_3 (\mu_0)$  is proportional to $ (m_{b} / M_W \cos\beta )^2$. \cite{demir}
However, the CKP edm constraint puts a strong constraint for large 
$\tan\beta$ case, which was not properly included in  Ref.~\cite{demir}.  
In Fig.~\ref{fig1} (a),
we plot $ 2 \theta_d \equiv {\rm Arg}~(M_{12}^{\rm FULL} / 
M_{12}^{\rm SM} )$ as a function of $\tan\beta$.
%
The open squares (the 
crosses) denote those which (don't) satisfy the CKP edm constraints. 
It is clear that the CKP edm constraint on $2 \theta_d$ is in fact very 
important  for large $\tan\beta$, and we have 
$| 2 \theta_d | \leq 1^{\circ}$. If we ignored the CKP edm constraint at 
all, then $|2 \theta_d |$ could be as large as $ \sim 4^{\circ}$.  
This observation is important for the CKM phenomenology, 
since time-dependent CP asymmetries in neutral $B$ decays into 
$J/\psi K_S, \pi\pi$ etc. would still measure directly three angles of the 
unitarity triangle even in the presence of new CP violating phases, 
$\phi_{A_t}$ and $\phi_{\mu}$.  Our result is at variance with that obtained
in Ref.~\cite{demir} where CKP edm constraint was not properly included.

\subsection{Dilepton Asymmetry}
If we parametrize the relative ratio of $M_{\rm SM}$ and $M_{\rm SUSY}$ as
$M_{\rm SUSY} / M_{\rm SM} = h e^{-i \theta}$, the dilepton asymmetry is 
given by 
\begin{equation}
A_{ll} = \left( {\Delta \Gamma \over \Delta M } \right)_{\rm SM}
f(h,\theta) \equiv 4~ {\rm Re} (\epsilon_B),
\end{equation}
where $f(h,\theta) = h \sin\theta/(1+2h\cos\theta + h^2)$ and 
$( \Delta \Gamma / \Delta M )_{\rm SM} = (1.3 \pm 0.2 ) \times 10^{-2}$.  
We have neglected the small SM contribution. It is about 
$\sim 10^{-3}$ in the quark level calculation, \cite{acuto} 
but may be as large as $\sim 1 \%$ if the delicate cancellation between the 
$u$ and $c$ quark contribution is not achieved. \cite{wolf}
The result of scanning over the available MSSM parameter space is that
$| f(h,\theta) | \leq 0.1$ so that $| A_{ll} | \leq 0.1 \%$, which is well 
below the current data, $A_{ll} = (0.8 \pm 2.8 \pm 1.2 ) \%$. \cite{opal}
On the other hand, if any appreciable amount of the dilepton asymmetry is 
observed, it would indicate some new CPV phases in the off-diagonal 
down-squark mass matrix elements, \cite{randall} 
assuming the MSSM is realized in nature. 

\subsection{$\Delta M_B$} 
On the contrary to the $\theta_d$ and $A_{ll}$ discussed in the previous 
paragraphs, the magnitude of $M_{12}$ is related with the mass difference of 
the mass eigenstates of the neutral $B$ mesons : 
$\Delta m_B = 2 | M_{12}| = (3.05 \pm 0.12) \times 10^{-13}~{\rm GeV}$, 
and thus it will affect the determination of $V_{td}$ from the $B^0 - 
\overline{B^0}$ mixing. We have considered $|M_{12}^{\rm FULL} 
/M_{12}^{\rm SM}|$ and its correlation with 
${\rm Br}(B\rightarrow X_s \gamma)$  are shown in Fig.~\ref{fig1} (b).
The deviation from the SM can be as  large as $\sim 60 \%$, and the 
correlation behaves differently from the minimal supergravity case. 
\cite{goto}
We repeated the same analyses for $B_s^0 - \overline{B_s^0}$ mixing. There is
no large new phase shift ($2 |\theta_s|$) in this case either, but the modulus 
of $M_{12} (B_s)$ can be enhanced by upto $60\%$ compared to the SM value.  
  
\section{Direct Asymmetry in $B \rightarrow X_s \gamma$}
The radiative decay of $B$ mesons, $B\rightarrow X_s \gamma$, is
described by the effective Hamiltonian including (chromo)magnetic dipole
operators. Interference between $b\rightarrow s \gamma$ and $b\rightarrow 
s g$ (where the strong phase is generated by the charm loop via 
$b\rightarrow c\bar{c}s$ vertex) can induce direct CP violation in 
$B\rightarrow X_s \gamma$, \cite{KN}  which is given by 
\begin{eqnarray}
 A_{\rm CP}^{b\rightarrow s\gamma} &  \equiv & 
{ \Gamma ( B \rightarrow X_{\bar{s}} + \gamma ) 
- \Gamma ( \overline{B} \rightarrow X_s + \gamma ) \over
 \Gamma ( B \rightarrow X_{\bar{s}} + \gamma ) 
+ \Gamma ( \overline{B} \rightarrow X_s + \gamma ) }
\nonumber  
\\
& \simeq & {1 \over | C_7 |^2 }~\left\{ 1.23 
{\rm Im}~[C_2 C_7^* ] - 9.52 {\rm Im}
~[C_8 C_7^*] \right.
\nonumber 
\\ & & \left. ~~~~~~~
+ 0.10 {\rm Im}~[C_2 C_8^*] \right\} ~({\rm in }~ \%) , 
\end{eqnarray} 
adopting the notations in Ref.~\cite{KN}.
We have ignored the small contribution from the SM, and assumed that the 
minimal photon energy cut is given by $E_{\gamma} \geq m_B (1-\delta)/2$ 
($\approx 1.8$ GeV with $\delta = 0.3$). $A_{\rm CP}^{b\rightarrow s \gamma}$
is not sensitive to possible long distance contributions
and constitute a sensitive probe of new physics that appears in the short 
distance Wilson coefficients $C_{7,8}$. \cite{KN} 

\begin{figure}[t]
    \begin{center}
      \begin{picture}(140,200)
      \put(-98,0){\psfig{file=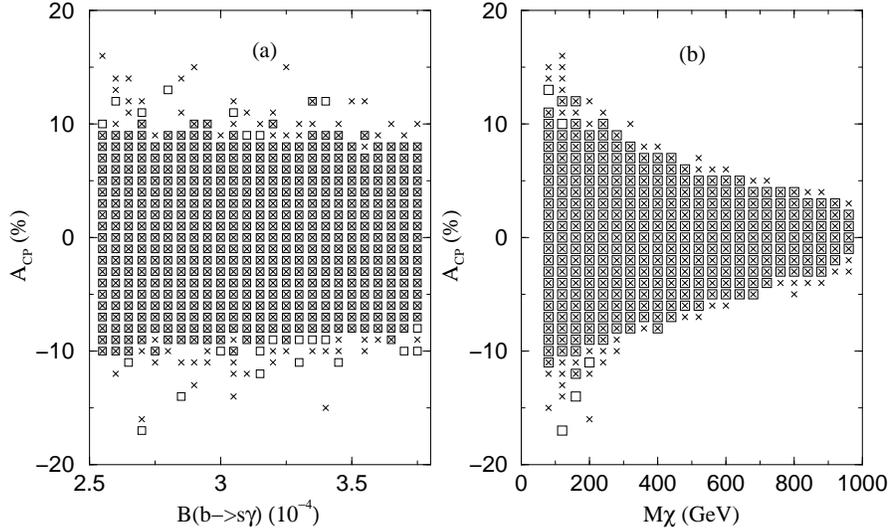,width=11.8cm}}
        \end{picture}
     \end{center}
\caption{
Correlations of $A_{\rm CP}^{b\rightarrow s\gamma}$ with 
(a) ${\rm Br} (B \rightarrow X_s \gamma)$ and (b) the lighter chargino
mass $M_{\chi^{\pm}}$.
The squares (the crosses) denote those which (do not) satisfy the CKP edm 
constraints. 
}
\label{fig2}
\end{figure}

The Wilson coefficients $C_{7,8}$ in the MSSM have been calculated by many
groups, \cite{bsgamma} including the PQCD corrections in certain MSSM 
parameter space \cite{giudice}. In this letter, we use the leading order 
expressions for $C_i$'s which is sufficient for $A_{\rm CP}^{b\rightarrow 
s\gamma}$.  After scanning over the MSSM parameter space described in 
Eq.~(3), we find that $A_{\rm CP}^{b\rightarrow s \gamma}$ can be as large 
as $\simeq \pm 16\%$ if chargino is light enough, even if we impose the edm 
constraints. Its correlation with $Br (B\rightarrow X_s \gamma)$ and chargino 
mass are shown in Figs.~\ref{fig2} (a) and (b) respectively.  
Our results are 
quantitatively different from other recent works, \cite{strumia},\cite{keum} 
mainly due to the different treatments of soft terms.
In the minimal supergravity scenario, this asymmetry is very small, because 
the $A_t$ phase effect is very small in the electroweak scale. \cite{keum}
If the universality assumption is relaxed, one can accomodate larger direct 
asymmetry without conflicting with the edm constraints.

\section{The Branching Ratio for $B \rightarrow X_s l^+ l^-$}
\begin{figure}
    \mbox{\psfig{width=5.9cm,figure=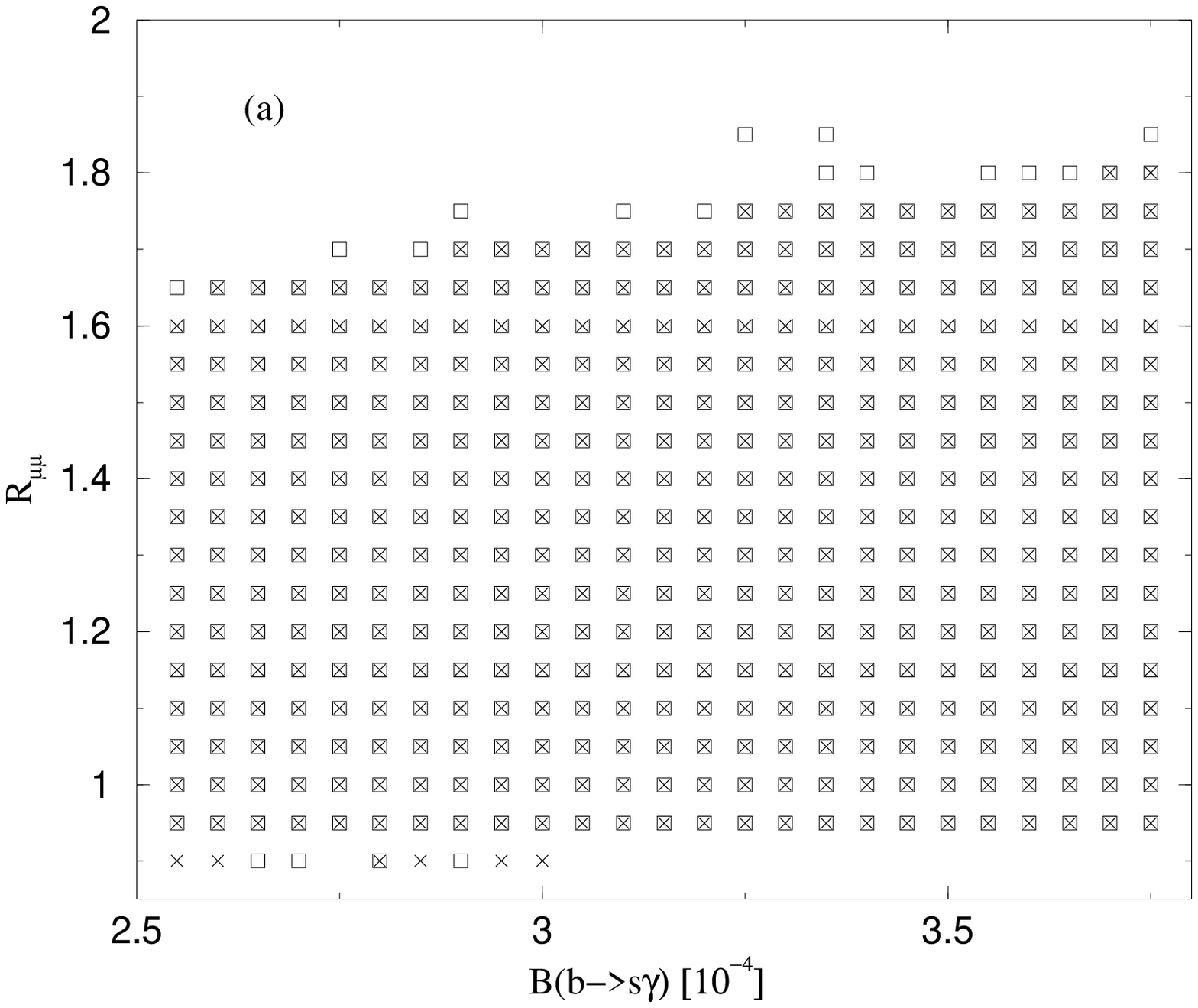}
          \psfig{width=5.9cm,figure=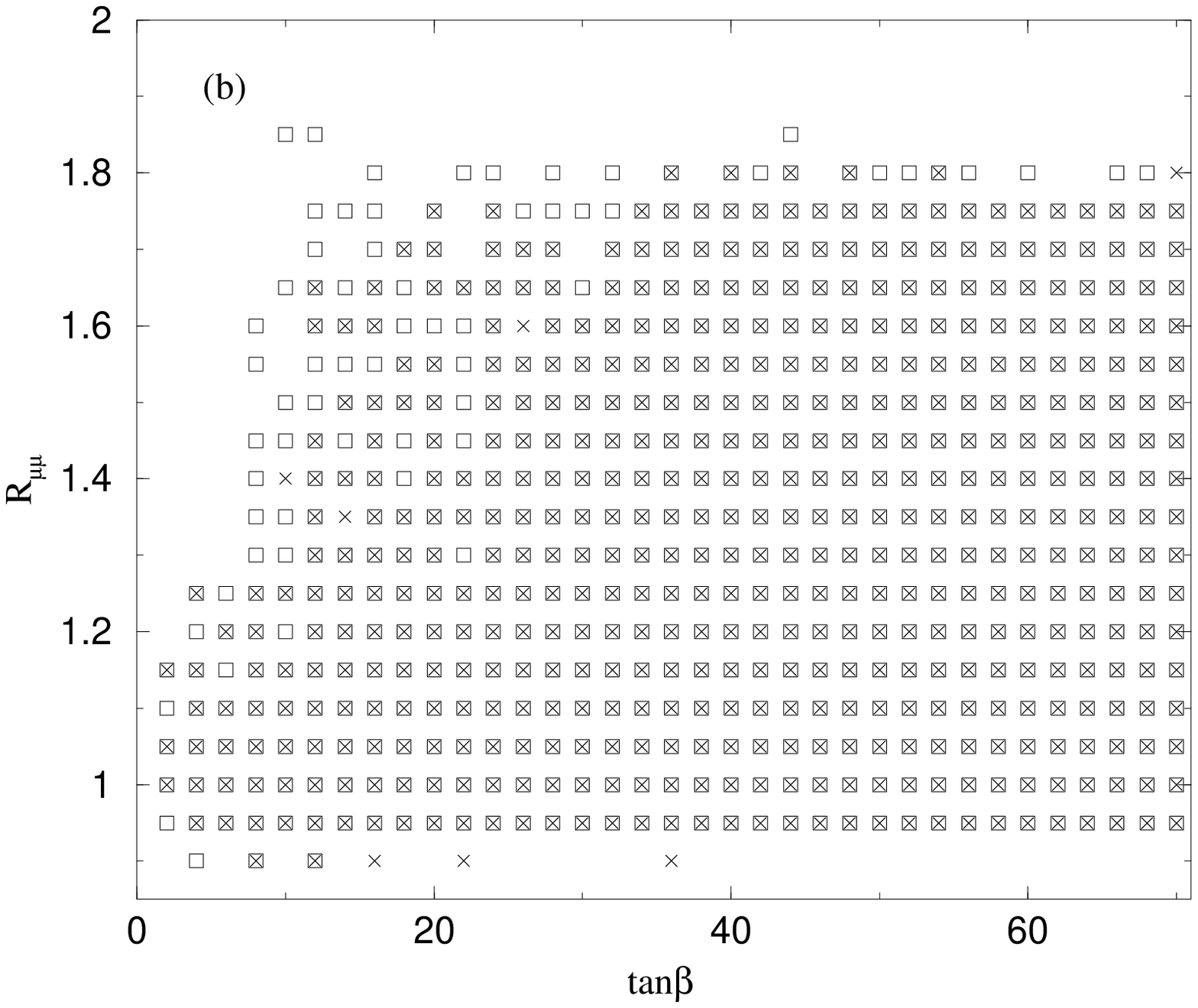}}
\vspace*{5mm}
\caption{
The correlations of $R_{\mu\mu}$ with ${\rm Br}(B\rightarrow X_s \gamma)$. 
The squares (the crosses) denote those which (do not) satisfy the CKP edm 
constraints. 
}
\label{fig3}
\end{figure}

Let us first consider the branching ratio for $B\rightarrow X_s l^+ l^-$. 
The SM and the MSSM contributions to this decay were considered by several 
groups. \cite{bsll},\cite{bsll_susy}    We use the standard 
notation for the effective Hamiltonian for this decay as described in Refs.~
\cite{bsll} and \cite{bsll_susy}. The new CPV phases 
in $C_{7,9,10}$ can affect the branching ratio and other observables in 
$B\rightarrow X_s l^+ l^-$ as discussed in the first half of Ref.~\cite{kkl}
in a model independent way. In the second half of Ref.~\cite{kkl}, specific 
supersymmetric models were presented where new CPV phases reside in flavor 
changing squark mass matrices.  In the present work, new CPV phases lie in 
flavor conserving sector, namely in $A_t$ and $\mu$ parameters. 
Although these new phases are flavor conserving, they affect the branching  
ratio of $B\rightarrow X_s l^+ l^-$ and its correlation with $Br (B\rightarrow 
X_s \gamma)$, as discussed in the first half of Ref.~\cite{kkl}.  Note that
$C_{9,10}$ depend on the sneutrino mass, and we have scanned  over 
$ 60 ~{\rm GeV} < m_{\tilde{\nu}} < 200~{\rm GeV}$.
In the numerical evaluation for $R_{ll} \equiv {\rm Br} (B\rightarrow X_s 
l^+ l^-) / {\rm Br} (B\rightarrow X_s l^+ l^-)_{\rm SM}$, we considered 
the nonresonant contributions only for simplicity, neglecting the 
contributions  from $J/\psi, \psi^{'}, etc$.. 
It would be straightforward to incorporate these resonance effects.
In Figs.~\ref{fig3} (a) and (b),
we plot the correlations of $R_{\mu\mu}$ with  
${\rm Br} (B\rightarrow X_s \gamma)$ and $\tan\beta$, respectively.
Those points that (do not) satisfy the CKP edm constraints are denoted by 
the squares (crosses). Some points are denoted by both the square and the 
cross. This means that there are two classes of points in the MSSM parameter 
space, and for one class the CKP edm constraints are satisfied but for 
another class the CKP edm constraints are not satisfied, and these two 
classes happen to lead to the same branching ratios for $B\rightarrow X_s 
\gamma$ and $R_{ll}$.  In the presence of the new phases $\phi_{\mu}$ and 
$\phi_{A_t}$, $R_{\mu\mu}$ can be as large as 1.85, and the deviations from 
the SM prediction can be large, if $\tan\beta > 8$.   
As noticed in Ref.~\cite{kkl}, the correlation between the ${\rm Br}
( B\rightarrow X_s \gamma)$ and $R_{ll}$ is distinctly different from that 
in the minimal supergravisty case. \cite{okada} In the latter case, only the 
envelop of Fig.~\ref{fig3} (a) is allowed,
whereas everywhere in between is allowed 
in the presence of new CPV phases in the MSSM.  Even if one introduces the 
phases of $\mu$ and $A_0$ at GUT scale in the minimal supergravity scenario, 
this correlation does not change very much from the case of the minimal 
supergravity scenario with real $\mu$ and $A_0$, since the $A_0$ phase becomes 
very small at the electroweak scale because of the renormalization effects. 
\cite{keum}  Only $\mu$ phase can affect the  electroweak scale physics, but 
this phase is strongly constrained by the usual edm constraints so that 
$\mu$ should be essentially real parameter. 
Therefore the correlation between $B\rightarrow X_s \gamma$ and $R_{ll}$
can be a clean distinction between the minimal supergravity scenario and 
our model (or some other models with new CPV phases in the flavor changing 
\cite{kkl}).   

\section{$\epsilon_K$ in Our Model}
\subsection{$K^0 - \overline{K^0}$ Mixing}
The new complex phases in $\mu$ and $A_t$ will also affect the $K^0 - 
\overline{K^0}$ mixing. The relevant $\Delta S = 2 $ effective Hamiltonian 
is given by 
\begin{equation}
H_{\rm eff}^{\Delta S = 2} = - {G_F^2 M_W^2 \over (2 \pi )^2}~
\sum_{i=1}^3 C_i Q_i,
\end{equation}
where 
\begin{eqnarray}
C_1 ( \mu_0 ) & = & \left( V_{td}^* V_{ts} \right)^2 \left[ 
F_V^W (3;3) + F_V^H (3;3) + A_V^C \right] 
\nonumber  \\
& + & \left( V_{cd}^* V_{cs} \right)^2 \left[ 
F_V^W (2;2) + F_V^H (2;2) \right] 
\nonumber  \\
& + & 2 \left( V_{td}^* V_{ts} V_{cd}^* V_{cs} \right) 
~\left[ F_V^W (3;2) + F_V^H (3;2) \right],
\nonumber  \\
C_2 ( \mu_0 ) & = & \left( V_{td}^* V_{ts} \right)^2 
F_S^H (3;3)  
 + \left( V_{cd}^* V_{cs} \right)^2 ~F_S^H (2;2) 
\nonumber  \\
& + & 2 \left( V_{td}^* V_{ts} V_{cd}^* V_{cs} \right) ~F_S^H (3;2), 
\nonumber  \\
C_3 ( \mu_0 ) & = & \left( V_{td}^* V_{ts} \right)^2 A_S^C,
\end{eqnarray}
where the charm quark contributions have been kept.
$G^{(3,k)i}$ and $H^{(3,k)i}$ are the same as Eqs.~(7) except that 
$m_b$ should be replaced by $m_s$ in the $K^0 - \overline{K^0}$ mixing.
Note that  $C_2 ( \mu_0 )$ was misidentified as $C_3^H ( \mu_0 )$ in 
Ref.~\cite{demir}.
The gluino and neutralino contributions are negligible in our model.
The Wilson coefficients at lower scales are obtained by renomalization
group running. The relevant formulae with the NLO QCD corrections at 
$\mu = 2$ GeV are given in  Ref.~\cite{contino}.  As in the 
$B^0 - \overline{B^0}$ mixing before,  $C_1 ( \mu_0 )$ and $C_2 ( \mu_0 )$ 
are real relative to the SM contribution in our model. On the other hand, 
the chargino exchange contributions to $C_3 (\mu_0 )$ (namely $A_S^C $) are 
generically complex relative to the SM contributions, and  can generate 
a new phase shift in the $K^0 - \overline{K^0}$ mixing relative to the 
SM value. This effect is in fact significant for large 
$\tan\beta (\simeq 1/\cos\beta)$, \cite{demir} 
since $C_3 (\mu_0)$  is proportional to $ (m_{s} / M_W \cos\beta )^2$. 

\begin{figure}
  \mbox{\psfig{figure=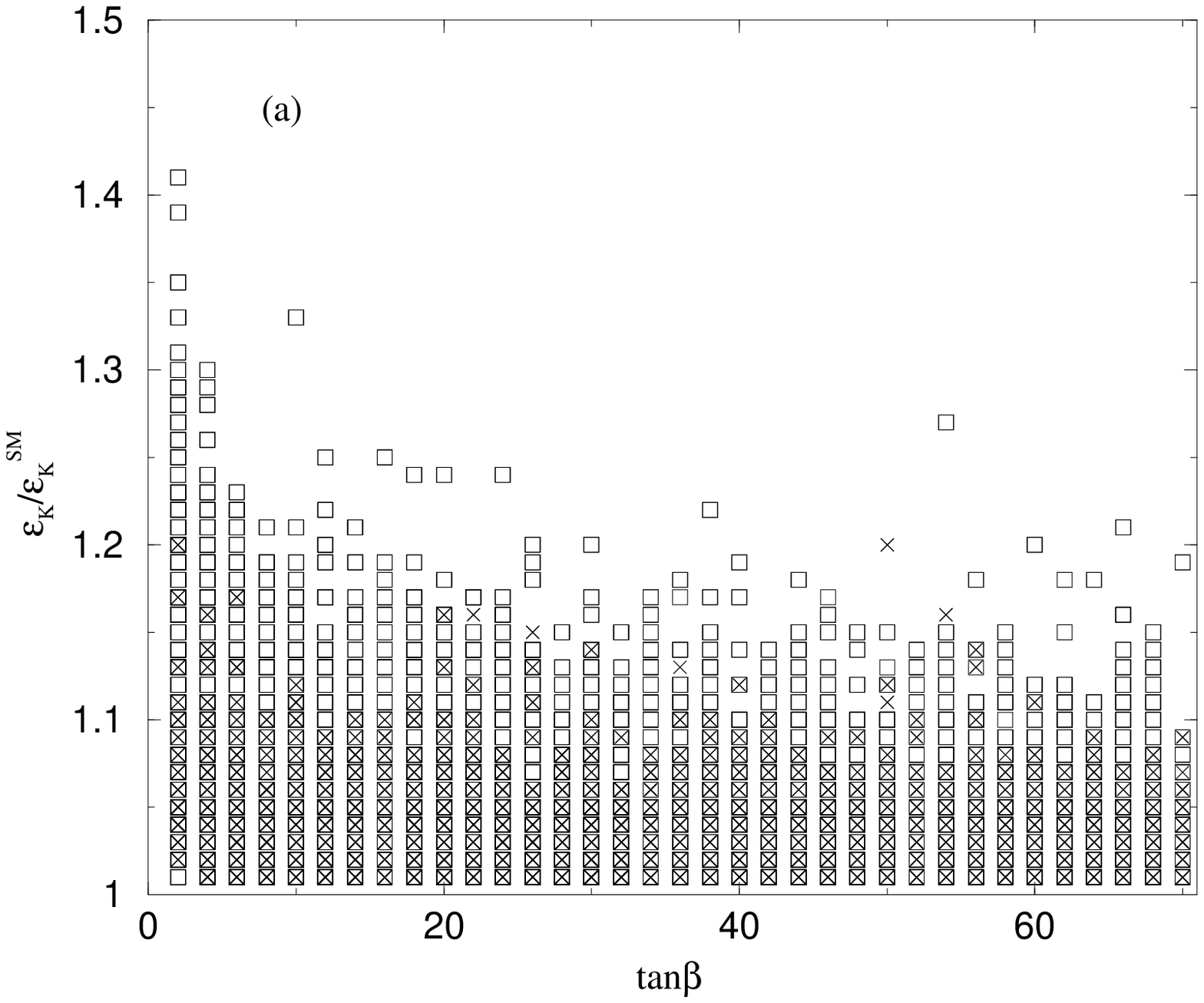,width=5.9cm}
        \psfig{figure=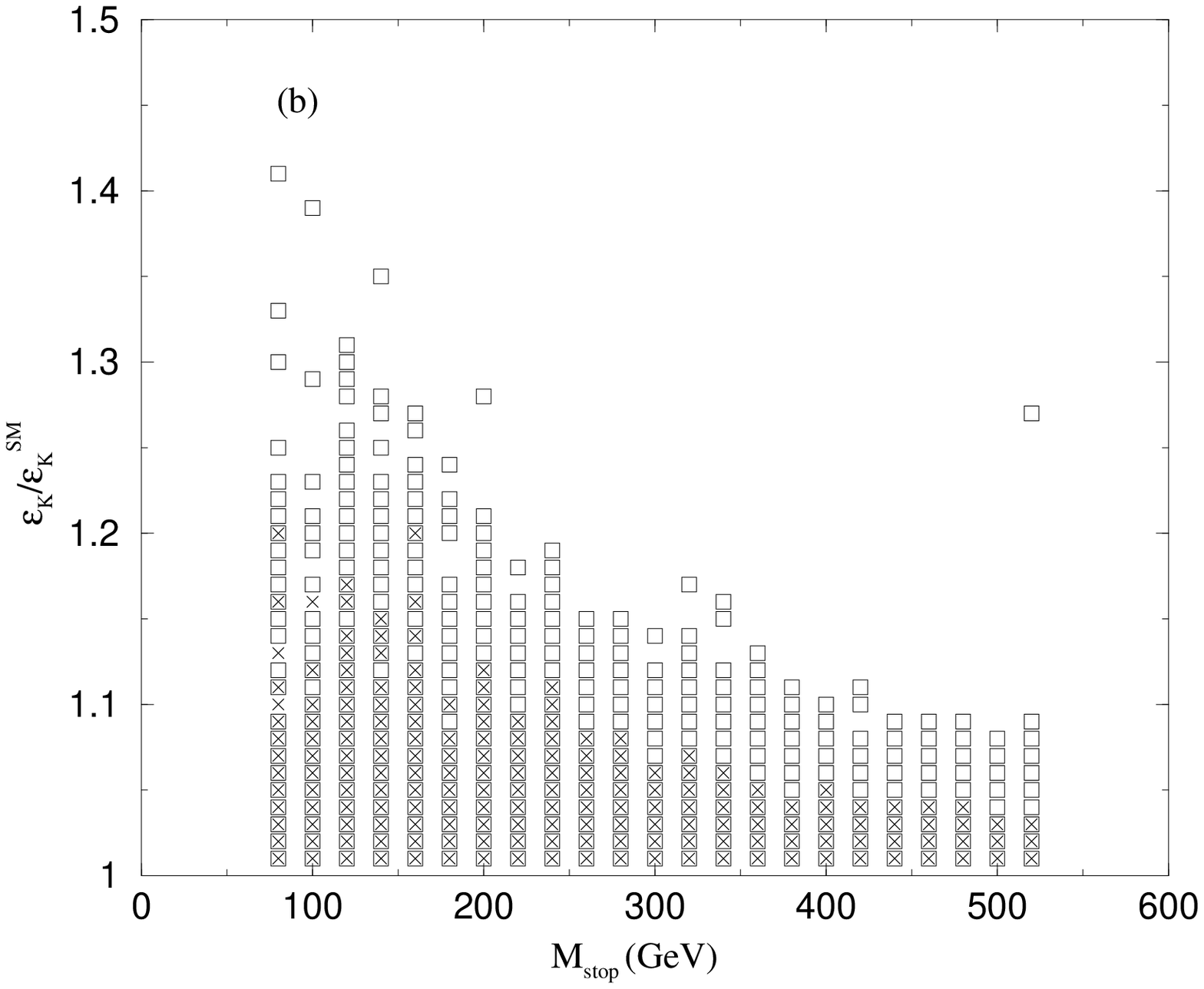,width=5.9cm}}
\vspace*{5mm}
\caption{
The correlations between $\epsilon_K / \epsilon_K^{SM}$ and the lighter 
chargino mass $M_{\tilde{\chi_1^{\pm}}}$ for (a) $2< \tan\beta < 35$  and 
(b) $35 < \tan\beta < 70$, respectively. 
The squares (the crosses) denote those which (do not) satisfy the CKP edm 
constraints. 
}
\label{fig4}
\end{figure}

The CP violating parameter  $\epsilon_K$ can be calculated from 
\begin{equation}
\epsilon_K \simeq {e^{i \pi / 4}~{\rm Im} M_{12} \over \sqrt{2} \Delta M_K},
\end{equation}
where $M_{12}$ can be obtained from the $\Delta S = 2 $ effective Hamiltonian 
through $2 M_K M_{12} = \langle K^0 | H_{\rm eff}^{\Delta S = 2} | 
\overline{K^0} \rangle$.  For $\Delta M_K$, we use the experimental value 
$\Delta M_K = (3.489 \pm 0.009) \times 10^{-12}$ MeV, instead of theoretical 
relation $\Delta M_K =  2 {\rm Re} M_{12}$,  since the long distance 
contributions to $M_{12}$ is hard to calculate reliably unlike the 
$\Delta S = 2$ box diagrams.  For the strange quark mass, we use the 
$\overline{\rm MS}$ mass at $\mu = 2$ GeV scale : 
$m_s (\mu = 2 {\rm GeV}) = 125$  MeV.
In Figs.~\ref{fig4} (a) and (b), we plot 
the results of scanning the MSSM parameter space : 
the correlations between $\epsilon_K / \epsilon_K^{\rm SM}$ 
and (a) $\tan\beta$ and (b) the lighter stop mass. 
We note that $\epsilon_K / 
\epsilon_K^{\rm SM}$ can be as large as $1.4 $  for $\delta_{KM} = 
90^{\circ}$ if $\tan\beta$ is small. This is a factor 2 
larger deviation from the SM compared to the minimal supergravity case. 
\cite{goto} The dependence on the lighter stop is close to the case of the 
minimal supergravity case, but we can have a larger deviations. 
Such deviation is reasonably close to the experimental value, and will 
affect the CKM phenomenology at a certain level. 

\subsection{Can $epsilon_K$ Come Entirely from $A_t$ and $\mu$ Phases ?}
In the MSSM with new CPV phases, there is an intriguing possibility that 
the observed CP violation in $K_L \rightarrow \pi\pi$ is fully due to the 
complex parameters $\mu$ and $A_t$ in the soft SUSY breaking terms which 
also break CP softly. 
This possibility was recently considered by Demir {\it et al.}. \cite{demir}  
Their claim was that it was possible to generate $\epsilon_K$ entirely from 
SUSY CPV phases for large $\tan\beta \approx 60$ with certain choice of
soft parameters. (Their choice of parameters leads to $M_{\chi^{\pm}} = 80$ 
GeV and $M_{\tilde{t}} = 85$ GeV, which are very close to the recent lower 
limits set by LEP2  experiments.) 
In such a scenario, only ${\rm Im}~(A_S^C)$ in Eq. (11) can contribute to 
$\epsilon_K$, if we ignore a possible mixing between $C_2$ and $C_3$ under 
QCD renormalization. In actual numerical analysis we have included this 
effect using the results in Ref.~\cite{contino}.  We repeated their 
calculations using the same set of parameters, but could  not confirm their 
claim. For $\delta_{KM} = 0^{\circ}$, we found that the supersymmetric 
$\epsilon_K$ is less than $\sim 2\times 10^{-5}$, which is too small compared 
to the observed value : $ | \epsilon_K | = (2.280 \pm 0.019) \times 10^{-3}$ 
determined from  $K_{L,S} \rightarrow \pi^+ \pi^-$. \cite{pdg98}

Let us give a simple estimate for supersymmetric $\epsilon_K$ with real CKM 
matrix elements, in which case only $C_3 ( \mu_0 )$ develops imaginary part 
and can contribute to $\epsilon_K$. For 
$m_{\tilde{t}_1} \sim m_{\chi^{\pm}} \sim M_W$, we would get 
$Y_2 \sim Y_2 (1,1,1,1) = 1/6$, and 
\[
| G^{(3,k) i} | \leq O(1),  ~~~{\rm and}~~~ 
| H^{(3,k) i} | \sim {m_s \tan\beta \over M_W}, 
\]
because any components of unitary matrices $C_R$ and $S_t$ are 
$\leq O(1)$. Therefore 
${\rm Im} ( A_S^C ) \leq O( 10^{-3} )$. 
Now using
\begin{equation}
{\rm Im} (M_{12}) = -{G_F^2 M_W^2 \over (2 \pi )^2} 
f_K^2 M_K 
\left( {M_K \over m_s } \right)^2~{1\over 24}~B_3(\mu)~{\rm Im} (C_3(\mu)),
\end{equation}
and Eq. (9), we get $|\epsilon_K | \leq 2 \times 10^{-5}$.

\section{Conclusion}
In conclusion, we assumed that the MSSM has new CPV phases beyond the CKM 
phase, and considered its observable consequences at $B$ factories and on
$\epsilon_K$ without making universality assumption on the soft terms at 
the GUT scale. Our study includes the EW baryogenesis scenario in the MSSM.
The main results can be summarized as follows. 
\begin{itemize}
\item 
There is no appreciable new phase in the $B^0-\overline{B^0}$ mixing
$( | 2 \theta_d | \leq 1^{\circ}$), so that time-dependent  
CP asymmetries in neutral $B$ decays (into $J/\psi K_S, \pi\pi$ etc.) 
still measure essentially three angles of the unitarity triangle even if 
there are new complex phases in $\mu$ and $A_t$ parameters.
\item 
The size of the $B^0-\overline{B^0}$ mixing can be enhanced up to 
$\sim 60 \%$ compared to the SM contribution, which will affect 
determination of $V_{td}$ from  $\Delta m_{B}$. 
\item 
There is no large shift in ${\rm Re} ( \epsilon_B )$, and 
dilepton CP asymmetry is rather small ($| A_{ll} | \leq 0.1 \%$).
\item 
Direct CP asymmetry in $B\rightarrow X_s \gamma$ can be as large as 
$\sim \pm 16 \%$ if chargino is light enough. This would encompass the 
interesting EWBGEN scenario in the MSSM. 
\item The branching ratio for $B\rightarrow X_s l^+ l^-$ can be enhanced upto 
$\sim 85 \%$ compared to the SM prediction, and the correlation between 
${\rm Br} (B\rightarrow X_s \gamma)$ and 
${\rm Br} (B\rightarrow X_s l^+ l^-)$ is distinctly different from the 
minimal supergravity scenario (CMSSM) (even with new CP violating phases)
\cite{okada} in the presence of new CP violating phases in $C_{7,8,9}$
as demonstared in model-independent analysis by Kim, Ko and Lee. \cite{kkl}
\item $\epsilon_K / \epsilon_{K}^{SM}$ can be as large as 1.4 for
$\delta_{KM} = 90^{\circ} $. This is the extent to which the new 
phases in $\mu$ and $A_t$ can affect the construction of the unitarity 
triangle through $\epsilon_K$. 
\item Fully supersymmetric CP violation is not possible even for large 
$\tan\beta \sim 60$ and light enough chargino and stop, contrary to the 
claim made in Ref.~\cite{demir}.  With real CKM matrix elements, 
we get very small $|\epsilon_K| \leq O(10^{-5})$, 
which is two orders of magnitude smaller than the experimental value.
\end{itemize}

These results would set the level of experimental sensitivity that one has to 
achieve in order to probe the SUSY-induced CP violations at $B$ factories 
through $B^0-\overline{B^0}$ and $A_{\rm CP}^{b\rightarrow s\gamma}$ mixing.  
Our results are conservative in a sense that we did not impose any conditions
on the soft SUSY breaking terms except that the resulting mass spectra for
chargino, stop and other sparticles satisfy the current lower bounds from 
LEP and Tevatron. 
Therefore, one would be able to find the effects of the phases of $\mu$ and 
$A_t$ parameters by observing $A_{\rm CP}^{b\rightarrow s\gamma}$ at 
B factories. 
Within our assumption,  the results presented here \cite{ko1}, \cite{ko2}
are conservative since we did not impose any conditions on the soft SUSY  
breaking terms except that the resulting mass spectra for chargino, stop
and other sparticles satisfy the current lower bounds from LEP and Tevatron.

Before closing this paper, we'd like to emphasize that all of our results 
are based on the assumption that there are no new CPV phases in the flavor 
changing sector. Once this assumption is relaxed, then 
gluino-mediated FCNC with additional new CPV phases may play 
important roles,  and many of our results may change. \cite{kkl}  For 
example, recently we have shown that both $\epsilon_K$ and $\epsilon^{'} / 
\epsilon_K$ can be saturated by a single complex parameter 
$(\delta_{12}^d )_{LL} \sim O(10^{-3})$ with $O(1)$ phase in the mass 
insertion approximation in supersymmetric models, if $| \mu \tan\beta| 
\sim O(10-20)$ TeV. \cite{ko3} In this case, $\epsilon_K$ is saturated by 
$(\delta_{12}^d )_{LL}$, whereas $\epsilon^{'} / \epsilon_K$ is given by 
the induced $(\delta_{12}^d )_{LR}^{\rm ind} \approx (\delta_{12}^d )_{LL}
\times ( m_s ( A_s - \mu \tan\beta) / \tilde{m^2} ) \sim 10^{-5}$. 
Remarkably, this can be achived without any contradiction to various FCNC or
EDM constraints.

%
%


\section*{Acknowledgments}
The authors thank A. Ali, G.C. Cho, J. Cline, A. Pilaftsis and O. Vives for 
useful communications. 
This work is supported in part 
by Korea Rsearch Foundation Program 1998-015-D00054 and the 
Distinguished Scholar Exchange Program of Korea Research Foundation (PK), 
and by KOSEF Postdoctoral Fellowship Program (SB).


\end{document}